\newif\iffinal
  \finaltrue

\documentclass[conference]{IEEEtran}
\usepackage{graphicx}
\usepackage{placeins}
\usepackage{float}
\usepackage{subfigure}

\usepackage{cite}
\usepackage{amsmath}
\usepackage{algorithm}
\usepackage{algpseudocode}
\usepackage{caption}
\usepackage{url}

\makeatletter
\let\OldStatex\Statex
\renewcommand{\Statex}[1][3]{%
  \setlength\@tempdima{\algorithmicindent}%
  \OldStatex\hskip\dimexpr#1\@tempdima\relax}
\makeatother

\iffinal
  \usepackage[finalold]{trackchanges} % also use \sr,\sj,\ak,\ba
\else
  \usepackage[inline]{trackchanges}
\fi

\addeditor{$A\!K$}
\addeditor{$R\!B$}
\newcommand{\ak}[1]{\note[$A\!K$]{#1}}

\usepackage{stmaryrd}
\usepackage{amssymb}

\usepackage{wrapfig}

\usepackage{hyperref}   % goes last
\hypersetup{
    colorlinks,
    linkcolor=black,
    citecolor=black,
}
\newcommand{\true}{\mathsf{true}}

\newcommand{\satcall}{\textsc{Sat}}

\newcommand{\xvars}{\overline{x}}
\newcommand{\ivars}{\overline{i}}
\newcommand{\ovars}{\overline{o}}

\newcommand{\xvarsE}{\overline{x}^\centerdot}
\newcommand{\ovarsE}{\overline{o}^\centerdot}
\newcommand{\cvars}{\overline c}
\newcommand{\fvars}{\overline f}

\newcommand{\xvals}{\mathbf{\overline{x}}}

\newcommand{\ivals}{\mathbf{\overline i}}
\newcommand{\ovals}{\mathbf{\overline o}}

\newcommand{\cvals}{\mathbf{\overline c}}
\newcommand{\fvals}{\mathbf{\overline f}}

\newcommand\Terr{\ensuremath{T^{\text{\raisebox{1.75pt}{$\centerdot$}}}}}
\DeclareMathOperator\xor\oplus
\DeclareMathOperator\andeq{\,\land\text{=}~}

\newcommand\li{\begin{itemize}}
\newcommand\il{\end{itemize}}
\renewcommand{\-}{\item}

\renewcommand{\iff}{\leftrightarrow}

\definecolor{darkgreen}{rgb}{0,0,0.2}
\definecolor{darkblue}{rgb}{0,0,.5}
\definecolor{darkred}{rgb}{0.9,0,0}
\definecolor{pink}{rgb}{0.95,0.08,0.55}
\definecolor{mygray}{gray}{.3}
\definecolor{lightblue}{rgb}{0.4,0.7,0.7}

\newcommand{\specialcellC}[2][c]{%
  \begin{tabular}[#1]{@{}c@{}}#2\end{tabular}}

\begin{document}

\title{OpenSEA:\\ Semi-Formal Methods for Soft Error Analysis}

%%\author{
%%\IEEEauthorblockA{
%%Roderick Bloem\IEEEauthorrefmark{1},
%%Ayrat Khalimov\IEEEauthorrefmark{1},
%%Patrick Klampfl\IEEEauthorrefmark{1},
%%Robert K\"onighofer\IEEEauthorrefmark{1},
%%Aiman Abu-Yonis\IEEEauthorrefmark{2},
%%Shiri Moran\IEEEauthorrefmark{2} \\
%%\IEEEauthorrefmark{1}TU Graz, Austria\\
%%\IEEEauthorrefmark{2}IBM Research Lab, Haifa, Israel
%%}
%%}

\author{
\IEEEauthorblockA{
Patrick Klampfl,
Robert K\"onighofer,
Roderick Bloem,
Ayrat Khalimov (TU Graz, Austria) \\
Aiman Abu-Yonis, Shiri Moran (IBM Research Lab, Haifa, Israel)
}
}

\maketitle

\begin{abstract}
Alpha-particles and cosmic rays cause bit flips in chips.
Protection circuits ease the problem, but cost chip area and power,
and so designers try hard to optimize them.
This leads to bugs:
an undetected fault can bring miscalculations,
the checker that alarms about harmless faults incurs performance penalty.
Such bugs are hard to find:
circuit simulation with tests is inefficient since it enumerates
the huge fault time-location space,
and formal methods do not scale since they explore the whole inputs.
In this paper,
we use formal methods on designer's input tests,
while keeping time-location open.
This idea is at the core of the tool OpenSEA.
OpenSEA can
(i) find latches vulnerable to and protected against faults,
(ii) find tests that exhibit checker false alarms,
(iii) use fixed and open inputs, and
(iv) use environment assumptions.
Evaluation on a number of industrial designs
shows that OpenSEA produces valuable results.
\end{abstract}

%%%\begin{abstract}
%%%With shrinking component sizes and an increasing density,
%%%hardware did not just become more powerful.
%%%As a side effect,
%%%it suffers from an increasing susceptibility
%%%to faults caused by cosmic radiation and alpha particles.
%%%
%%%Today's critical systems rely on error-detecting or error-correcting codes.
%%%Verifying them by hand is a time-consuming and error-prone task,
%%%which is why we propose new methods that can be applied automatically. % TODO: propose/compare new/existing????
%%%
%%%First, methods to detect bugs in a protection logic are presented.
%%%They can be used to either find unprotected components
%%%or to test if a protection logic reports errors too often.
%%%This is achievable by thorough formal methods,
%%%by a fast but inaccurate simulation-based technique,
%%%or by a semi-formal approach that provides more flexibility.
%%%
%%%Afterwards, an approach to verify definitely protected components is shown.
%%%It formally proves components that are resilient in each situation
%%%by over-approximating the reachable state-space. 
%%%
%%%Our approaches do not assume any particular structure of the circuits and can be applied early in the design process.
%%%The experimental results show that our semi-formal approach already outperforms simulation if just a few input values are set open.
%%%In addition, experimental results show that the tool is applicable and produces valuable results on industrial designs.
%%%\end{abstract}

\maketitle

\section{Introduction}
In 1965 Moore's law was born~\cite{Moore1965}:
chips density doubles every few years.
Aside from good things like improving performance,
increasing chips density also increases sensitivity to
cosmic radiation and alpha-particles~\cite{rayeffect, 4336494, Ziegler}.
These external influences can cause bit flips
--- the sudden change of internal memory values ---
that can lead to hardware misbehaviours~\cite{Baumann:2005}.
Such errors are called \emph{soft errors},
because the hardware is not harmed permanently.

There are ways to protect the circuit against soft errors~\cite{serrmitigation, Siewiorek:1998, austin1999diva, bower2005}. 
\emph{Triple modular redundancy} copies the circuit three times,
computes three results in parallel,
and chooses the result by majority vote,
thus \emph{overcoming} soft errors.
\emph{Error correcting codes} (ECCs) introduce spare bits that are used to \emph{restore} the original values after the bit flip happens~\cite{hamming}.
Parity computations~\cite{austin} can \emph{detect} bit flips.

But such protection methods are too expensive,
since often circuits are intrinsically protected~\cite{Mukherjee:2003},
for example, when the computation will not be used
(think of branch predictions in processors).
Thus, designers carefully optimize the protection logic to save area and power, which is difficult and error-prone.

We present tool OpenSEA to help the designer
to verify and optimize the protection circuit.
We assume that the protection circuit has a special output ``alarm''
(risen when a fault is detected),
and consider only single fault scenarios~\cite{315626}.

OpenSEA provides three kinds of algorithms.
First, it searches for vulnerable latches.
Roughly,
a latch is vulnerable if the fault in it can escape to the user
without being alarmed.
Second, OpenSEA searches for protected latches.
Roughly,
a latch is protected if the fault in it is always alarmed or recovered from
before propagating to the user.
Third, OpenSEA searches for spurious alarms.
Roughly, an alarm is raised spuriously when the fault is always recovered from without being noticed.
Spurious alarms hurts performance.

The idea behind the algorithms is to compare a fault-free circuit with
a modified circuit where we model a fault, see Fig.~\ref{fig:mc_schema}.
For example,
if run a given test case on two circuits,
inserting a fault into the second circuit,
and check if the outputs can diverge without the alarm,
then we get, roughly, the simulation based algorithm.
Such two-circuits-comparison check can be offloaded to a SAT solver.
Then, we can keep the fault location and time ``symbolic'' (``free''),
thus asking a SAT solver if there is a fault latch and time that can escape without raising the alarm.
Such approach is not new and was explored in \cite{bmc_approx,DBLP:conf/fmcad/FrehseFAYD12,Krautz},
but to our knowledge no tools are available.
Also, we allow the designer to use fixed tests, partially fixed tests (with some inputs left open), or completely open tests.
Thus, our algorithm ranges from model checking to simulation.
We call such approach \emph{semi-formal}.

We evaluated OpenSEA on made-up examples, and started evaluation on industrial designs.

The source code and benchmark results are available at \url{https://extgit.iaik.tugraz.at/scos/soft-error-analysis/}.
%More algorithm details are available in thesis of Patrick Klampfl (in the repository).

\section{Preliminaries}

\subsection{Faults, Vulnerable and Protected Latches, Alarm}

A {\em (transient) fault}
is a nondeterministic change of the latch value for one time step.
We assume there can be at most one transient fault in a circuit during the execution.
We do not consider other types of faults (stuck-at-zero, byzantine, etc.).

We assume a circuit has a special output {\em alarm}:
$alarm=\true$ means a fault has been detected,
and a higher-level service handles it
(e.g., it restarts the computation).

Consider an execution where a fault happens in latch $x$ at moment $t$.
{\em The fault escapes}
if outputs diverge at some moment $t'$ and the alarm is not raised before or at $t'$.
{\em The fault is alarmed} if alarm is risen in the fault moment or later.
Note that a fault can escape and yet be alarmed (too late).

A {\em latch is vulnerable} if there is an execution with an escaping fault in the latch.
Otherwise, the {\em latch is protected}.
I.e.,
a latch is protected if on all executions,
a fault in the latch is either always detected
before or at the moment when the outputs diverge,
or the outputs never diverge.

Consider an execution with a fault at moment $t$ in latch $x$.
An {\em alarm is spurious at moment $\geq t$}
if the circuit states converge\ak{clarify} on all evolutions of that execution from $t$.

\subsection{Transition Relations}
\ak{execution, test case}
\begin{figure}[tb]
\center
\centering
\includegraphics[width=\columnwidth]{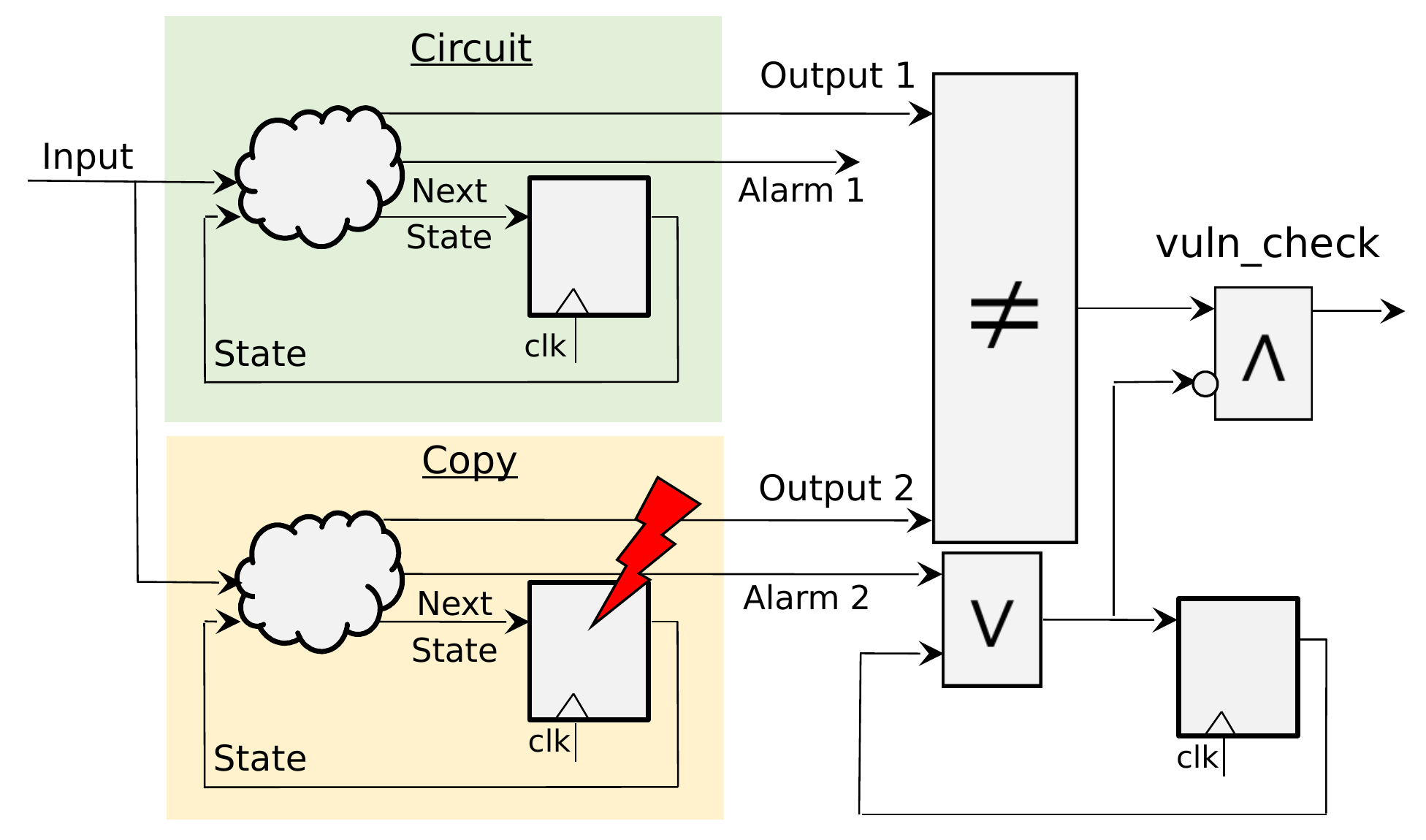}
\caption{Composed circuit used for vulnerability check}
\label{fig:mc_schema}
\vspace{-0.7cm}
\end{figure}

\noindent\textit{Notation.}
We assume the reader is familiar with the notions of
literal, clause, cube, CNF, DNF,
sequential circuit.
Tuples of Boolean variables are usually overlined,
e.g., $\xvars = (x_1, \ldots, x_n)$,
while tuples of their values
%(aka values)
are also bolded,
e.g., $\xvals = (1,0, ..., 1)$.
We assume $\bigwedge_{i}^{j} (..)$ is $\true$ when $j < i$.

Consider Fig.~\ref{fig:mc_schema}:
the top boxed area ``Circuit'' represents the circuit under test:
it is divided into latches (the rectangle) and the combinatorial part (the cloud),
it has inputs, outputs ``Output 1'',
and the special output ``Alarm 1''.

A circuit can be represented as a Boolean formula
$T(\xvars, \ivars, \ovars, a, \xvars')$ over variables $\xvars,\ivars,\ovars,a,\xvars'$,
called {\em transition relation}.
Given values $\xvals$, $\xvals'$, $\ivals$, $\ovals$, and $a$,
$T(\xvals, \ivals, \ovals, a, \xvals')$ is true
if the circuit, starting with latch values $\xvals$ and reading inputs $\ivals$,
outputs $\ovals$, alarms $a$,
and updates latch values to $\xvals'$.
Latch values are called {\em circuit state}.

To model the faults,
we introduce new inputs $\overline{c}$ (where $|\overline{c}| = |\xvars|$) and $f$,
and use
$\Terr(\xvars, \ivars, f, \overline{c}, \ovars, a, \xvars')$ defined as
$$
T(\xvars,\ivars,\ovars, a, \xvars')[x_i' \mapsfrom x'_{orig}] \land \bigwedge_i x'_i \iff x'_{orig} \xor (f \land c_i).
$$
Intuitively,
$\Terr$ flips the original $x'_i$ when $c_i$ and $f$ are set;
$f$ encodes whether or not to inject a fault,
and $c_i=\true$ means that the latch $i$ should be flipped.
The single-fault assumption is modelled later.

%%%An execution of a circuit over $n$ time steps is encoded by unrolling the transition relation for $n$ times:
%%%     $T(\xvars_0, \ivars_0, \ovars_0,\xvars_1) 
%%%\land T(\xvars_1, \ivars_1, \ovars_1,\xvars_2)
%%%\land ... 
%%%\land T(\xvars_n, \ivars_n, \ovars_n,\xvars_{n+1})$.

\section{Classification Algorithms}
In this Section we describe two algorithms used by OpenSEA.
The first one searches for vulnerable latches and spurious alarms:
it can accept (fixed or open) tests.
The second algorithm searches for protected latches.
Both algorithms are sound but incomplete:
the classification of the latches they produce is correct,
but some latches may be unclassified.

%%%%\subsection{More Notation}
%%%%$(\sat, \vvals) = \satcall(F(\vvars))$
%%%%denotes a call to a SAT solver.
%%%%If $F$ is satisfiable,
%%%%then $\sat$ is set to $\true$ and
%%%%$\vvals$ is the satisfying assignment.
%%%%If $F$ is unsatisfiable, then $\sat$ is set to $\false$.

%% \subsection{Simulation}
%% Although not being complete, simulation or emulation \cite{PellegriniCZSBA08, Goswami97depend} are widely used because of the good scalability.
%% This section presents our simulation based algorithm, which analyzes a circuit using \emph{test cases}.
%% Algorithm \ref{alg:sim_basic} compares simulation results of a fault-free circuit with simulations that contain an induced fault.
%% It is the foundation for two algorithm variants, one checks for vulnerable latches (\ref{alg:sim_vuln}), and one for false positives (\ref{alg:sim_fp}).

\subsection{Searching for Vulnerable Latches and Spurious Alarms}

Algorithm~\ref{alg:vuln} searches for vulnerable latches.
The input is a test case $(\ivals_1, \ivals_2, ..., \ivals_t)$ of length $t$,
the output is the set of vulnerable latches,
and $F$ is the main Boolean formula.
Conceptually, $F$ encodes signal ``vuln\_check'' on Fig.~\ref{fig:mc_schema}.
We initialize the latches to $0$,
then in Line~\ref{alg:vuln:cardA} and \ref{alg:vuln:cardB} add to $F$
the constraints ensuring the single-fault assumption.
The for-loop incrementally unrolls (Line~\ref{alg:vuln:unroll}) the transition relations of
the non-faulty and faulty circuits up to the length of the test,
while reading the test inputs.
Lines~\ref{alg:vuln:vuln-specific}--\ref{alg:vuln:vuln-specific-end}
are specific to vulnerable latches (we later change them for spurious alarms):
we encode that the outputs differ while the alarm is low,
and use a SAT solver to incrementally find all vulnerable latches
  (Lines~\ref{alg:vuln:while:start}--\ref{alg:vuln:while:end})).
$\satcall(F \land A)$ denotes a SAT call:
it returns $0$ if the formula is unsatisfiable,
otherwise it returns a satisfying assignment for variables
$\fvars$ and $\cvars$.
On algorithm termination, $vulnerable$ contains vulnerable latches.
\vspace{-0.05cm}
\begin{algorithm}
\begin{algorithmic}[1]
\Procedure{FindVulnerable}
          {test case $(\ivals_1, ..., \ivals_m$)}
  \State vulnerable := $\{\}$
  \State set $\xvars_1$ and $\xvarsE_1$ to $(0,...,0)$
  \State $F := \bigl((\sum_{i \in [n]} c_i) = 1\bigr)$ \label{alg:vuln:cardA}
  \For{$s = 1$ to $m$}
    \State $F \andeq \bigwedge_{j < s}
                  (f_s \rightarrow \neg f_j)$ \label{alg:vuln:cardB}

    \State $F \andeq T(\xvars_s, 
                     \ivals_s, 
                     \ovars_s,
                     a,
                     \xvars_{s+1}) \land \neg a \land $
    \Statex[4] \!$\Terr(\xvarsE_s,
                        \ivals_s,
                        f_s,
                        \overline{c},
                        \ovarsE_s,
                        a_s^\centerdot,
                        \xvarsE_{s+1})$ \label{alg:vuln:unroll}

    \State $F \andeq \neg a_s^\centerdot$ \label{alg:vuln:vuln-specific}

    \State $A := (\ovars_s \neq \ovarsE_s)$

    \While{$(\fvals,\cvals) := \satcall\bigl(F \land A\bigr)$} \label{alg:vuln:while:start}
	    \State vulnerable.add($i$) $\textit{with } \cvals[i]=1$ \Comment{$i$ is unique}
    	\State $F \andeq \lnot c_i$
    \EndWhile \label{alg:vuln:vuln-specific-end} \label{alg:vuln:while:end}
  \EndFor
\EndProcedure  
\end{algorithmic}
\caption{Searching for vulnerable latches} \label{alg:vuln}
\vspace{-0.1cm}
\end{algorithm}

To search for spurious alarms,
OpenSEA uses the modified Algorithm~\ref{alg:vuln}
where lines \ref{alg:vuln:vuln-specific}--\ref{alg:vuln:vuln-specific-end}
are replaced by
\begin{figure}[H]
\vspace{-0.1cm}
\begin{algorithmic}
\State $F \andeq (\ovars_s = \ovarsE_s)$  \Comment{outputs are equal}
\State $A := (\xvars_{s+1} = \xvarsE_{s+1}) \land \bigvee_{j \leq s} a_j^\centerdot$ \Comment{states converge, but alarm is raised}
\While{$(\fvals,\cvals) := \satcall\bigl( F \land A \bigr)$} 
    \State spurious.add($i$) with $\cvals[i]=1$ \Comment{$i$ is unique}
    \State $F \andeq \neg c_i$
\EndWhile
\end{algorithmic}
\vspace{-0.2cm}
\end{figure}
\noindent
On termination, the variable $spurious$ contains the latches
whose faults can lead to spurious alarms (initially empty).

\subsubsection{Implementation}
\begin{figure}
\centering
\includegraphics[width=0.5\textwidth]{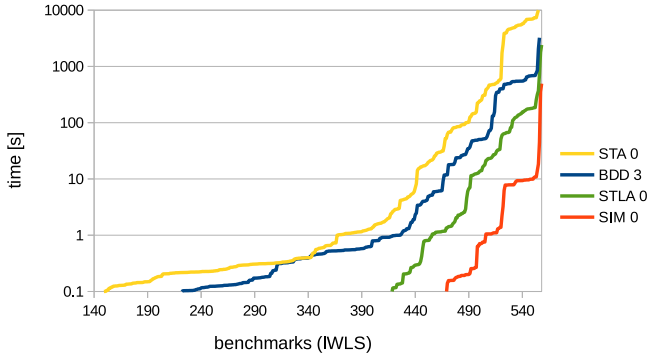}
\caption{Cactus plot. Detecting vulnerable latches: 3 concrete test cases, 15 time steps. We used a subset of IWLS'02 and IWLS'05 benchmarks, with added by us protection circuit. The benchmarks have up to 10k latches.}
\label{fig:exp:vuln:comparison:fixed}
\vspace{-0.5cm}
\end{figure}

We implemented these algorithms in OpenSEA.
The relevant to these checks tool's arguments are:
\li
\- Circuit is in the AIGER format \cite{aiger}.
   Optionally, you can provide the environment assumptions circuit,
   which purpose is to filter out traces not interesting for testing.
   %Also, you can provide a file listing latches that should be exclude from the checks.
\- Input test case is a file that contains input vectors.
   An inputs value can be left open by writing ``?''.
   % Alternatively,
   % the tool can generate random or symbolic tests.
\- OpenSEA supports several variations of the algorithms:
   \li
   \- \textsf{stla}: fault time and location are symbolic
      (Alg.~\ref{alg:vuln}),
   \- \textsf{sta}: only fault time is symbolic
      while latches are enumerated in the for-loop
   \- \textsf{sim} (vulnerability only):
       enumeration of fault time and location, 
       and of symbolic inputs
   \- \textsf{bdd} (vulnerability only):
      Algorithm~\ref{alg:vuln} using BDDs
   \il
\- OpenSEA outputs:
   vulnerable latches (or latches causing spurious alarms)
   and the witnessing traces
\il
Cactus plot in Fig.~\ref{fig:exp:vuln:comparison:fixed}
shows how the algorithm variations perform
on tests with no open inputs,
where, surprisingly, the simulation approach is the fastest
(in this experiment setup).
But the \textsf{stla} variation scales much better with
the increasing of open inputs:
freeing inputs turned out to have almost no effect on performance,
while it exponentially increases the simulation time.
In the extreme of all inputs open:
we compared \textsf{stla} with model checking approach,
where we created the circuit in Fig.~\ref{fig:mc_schema} and
passed it to a model checker (BLIMC) --- they perform comparably.
Also,
with the increasing length of test cases,
the simulation approach scales linearly,
while \textsf{stla} roughly exponentially.
Thus,
the designer can choose the appropriate algorithm variation 
%the vulnerability check algorithm
depending on the tests at hand.

As for spurious alarms algorithm variants,
\textsf{stla} and \textsf{sta},
the first version performed several orders faster.

\vspace{-0.1cm}
\subsection{Searching for protected latches}
A \emph{latch $i$ is 1-step protected} if the formula
\vspace{-0.1cm}
\begin{equation}
\begin{aligned}
      & ~T\big((x_1, ..., ~x_i, ..., x_n), \ivars, \ovars, a, \xvars'\big) \land \neg a \\
\land~ & T\big((x_1, ..., \neg x_i, ..., x_n), \ivars, \ovarsE, a^\centerdot, {\xvarsE}'\big) \\
\land~ & \neg a^\centerdot \land (\ovars \neq \ovarsE \lor \xvars' \neq {\xvarsE}')
\end{aligned}
\label{eq:1-protected}   
\vspace{-0.1cm}
\end{equation}
is unsatisfiable.
Such a check means:
no matter in which state the circuit is and no matter which inputs are used,
if latch $i$ flips its value,
the error is either already gone in the next time step and the outputs are equal,
or the alarm is raised.

Some error-detection and correction methods may take more than one clock cycle
to raise an alarm or to recover, making this check too conservative.
Hence we similarly define a \emph{$k$-protected latch}
by unrolling the transition relation for $k$ steps.

Another source of over-conservativeness is that the check
allows the circuit to start from any state, including unreachable.
We use formula $approx$ to restrict the starting states $\xvars_1$:
for $j\geq 0$,
\vspace{-0.2cm}
\begin{equation*} %\label{eq:j-approx}
\begin{aligned}
approx(j) := &\left( \bigwedge_{i=1}^{j} T(\xvars_{i-j}, \ivars_{i-j}, \ovars_{i-j}, a, \xvars_{(i-j)+1}) \land \neg a \right) \\
& ~\lor \xvars_1 = reach(\overline{0}, j)
\end{aligned}
\end{equation*}
where
$reach(\overline{0}, j)$ encodes all states
reachable from the initial state $\overline{0}$ within $j$ time steps,
the term in parentheses encodes all reachable states from any state after
exactly $j$ steps.
Thus,
in Eq.~\ref{eq:1-protected} we can replace the starting states $\xvars$ by $\xvars_1$ and
conjunct it with $approx(j)$.
This way we can define a \emph{$jk$-protected latch} where $j \geq 0, k \geq 1$.

Algorithm~\ref{alg:prot} finds all $jk$-protected latches.
It uses the above two ideas,
but instead of checking every latch,
it checks them all incrementally (this happened to be faster).
This algorithm starts by assuming that all latches are $jk$-protected.
Line~\ref{alg:prot:cardinality} encodes the single-fault assumption.
Lines~\ref{alg:prot:1-step:start}--\ref{alg:prot:1-step:end}
encode whether it is possible,
without raising the alarm,
to diverge the outputs within $k$ steps or the states at step $k+1$.
If yes, then the SAT solver returns latch $i$ which leads to divergent outputs or states,
and we exclude it from $protected$.
We repeat this until the formula becomes unsatisfiable,
whereupon declare the remaining latches $jk$-protected.
\begin{algorithm}
\begin{algorithmic}[1]
\Procedure{FindProtected}{$j\geq 0$, $k\geq 1$}
  \State protected := all latches
  \State $F := approx(j)$
  \State $F \andeq f_1 \land \bigwedge_{s=2}^k \neg f_s \land \bigl((\sum_{i \in [n]} c_i) = 1\bigr)$ \label{alg:prot:cardinality}
  \State $F \andeq \bigwedge_{s=1}^{k} T(\xvars_s,\ivars_s,\ovars_s, a_s,\xvars_{s+1}) \land \lnot a_s$ \label{alg:prot:unfold-orig} \label{alg:prot:1-step:start}
  \State $F \andeq \bigwedge_{s=1}^{k} \Terr(\xvarsE_s,\ivars_s,f_s,\cvars,\ovarsE_s,a_s^\centerdot,\xvarsE_{s+1})$ \label{alg:prot:unfold-faulty}
  \State $F \andeq (\neg a_1^\centerdot \land \ovars_1 \neq \ovarsE_1)$ \label{alg:prot:1-step:end}
\item[] \hspace{1.3cm}  
        $\lor (\neg a_1^\centerdot \land \neg a_2^\centerdot \land \ovars_2 \neq \ovarsE_2)$
\item[] \hspace{1.3cm}  $\lor \ldots$
\item[] \hspace{1.3cm}
   $\lor
   \bigl(\neg a_1^\centerdot \land \ldots \land \neg a_k^\centerdot \land
        (\ovars_k \neq \ovarsE_k \lor \xvars_{k+1} \neq \xvarsE_{k+1})
   \bigr)$ 
  \While{$\cvals := \satcall(F)$}
    \State protected.remove($i$) with $\cvals[i]=1$ \Comment{$i$ is unique}
    \State $F \andeq \neg c_i$
  \EndWhile
\EndProcedure  
\end{algorithmic}
\caption{Searching for Protected Latches} \label{alg:prot}
\end{algorithm}

\vspace{-0.2cm}
\subsubsection{Implementation}
For searching protected latches,
OpenSEA accepts a circuit, numbers $j$ and $k$.
It does not accept neither test cases nor an environment model.
We experimented with made-by-us triple module redundancy adder,
whose latches are $10$-protected (thus, $j=1$, $k=10$).
Search of $10$-protected latches took $\approx 100$ times longer
than that of $1$-protected latches.

\section{Experiments on Industrial designs}
\label{sec:IBM}
We present preliminary OpenSEA experiments on circuits from IBM microprocessor.
We start with some preliminaries definitions and then describe our experiments.

\begin{figure}
\centering
\includegraphics[width=1.0\linewidth]{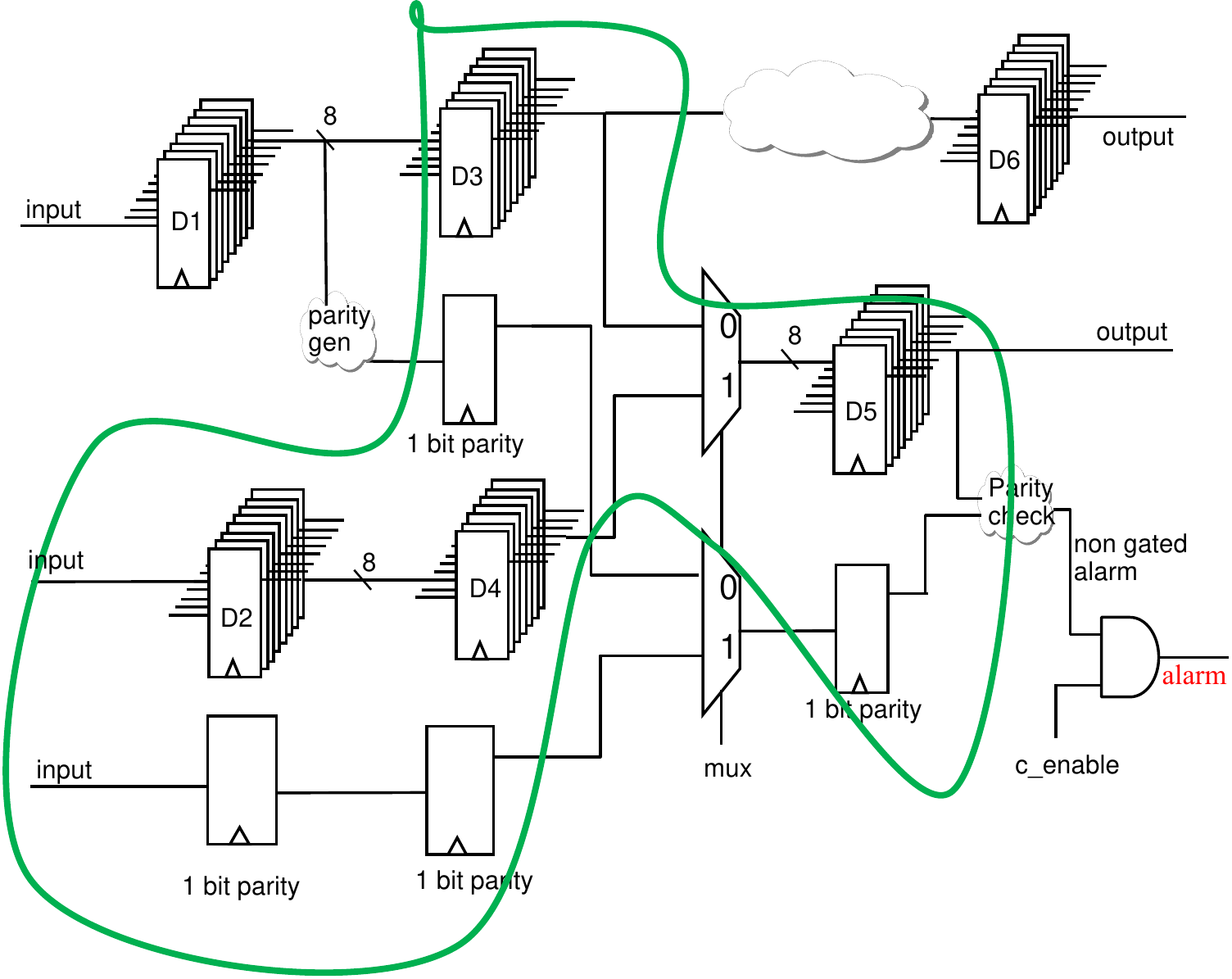}
\caption{Typical industrial circuit. Latches D1, D3, and D5 are parity-protected in this macro,
         while parity generation for D2 and D4 is done externally and the result is given as the bottom input.
         Signal $mux$ is used to choose the data path between D1-D3 and D2-D4,
         signal $c\_enable$ is used to disable (to \emph{gate}) the alarm.}
\label{fig:parity_protected_structure}
\vspace{-0.5cm}
\end{figure}

\emph{Macros}:
Large industrial designs are usually divided into sub-components, called {\em macros}, and 
in many cases the protection logic spans over more than one macro.
E.g., the parity check may reside in one macro while the parity generation resides in a neighboring macro.
For example, in Fig.~\ref{fig:parity_protected_structure}
the parity generation associated with data bus $D2$ is performed in a neighboring macro.
%It should be noted that due the capacity and to different responsibilities,
%in many cases the current scope consists of only a part of the design.

\emph{Gating}:
In high performance designs in majority of the cases a fault has no effect on the application~\cite{ArbelBHKKM16},
since the affected data will not be actually used.
Thus, to avoid unnecessary recoveries and also to save power,
circuits include dedicated logic to mask out the alarm signal when the relevant protected data is not used. 
For example, in Fig.~\ref{fig:parity_protected_structure} the signal  $c\_enable$ is used to disable the alarm when the 
output of $D5$ and of $D6$ do not effect the application.  We call such masking of the alarm signal \emph{gating}.
This introduces an additional verification challenge, since it is not sufficient to verify that a given sequential element is {\em protected}, 
it is also required to verify that it is not {\em over gated}---namely, that when ever it is gated it indeed does not affect the output.
In addition, it is desirable to also detect {\em spurious alarms}---cases in which the alarm rises even though the contaminated value will have no effect. 
%the signal driving
%a component with a protection logic and gating is only {\em potentially protected};
%it is protected only if there is no {\em over gating} that prevents the alarm from rising when the relevant data will propagate to the output. 
%For example, in Fig.~\ref{fig:parity_protected_structure}
%the encircled component is {\em potentially protected};
%data bus $D5$ is protected only if the c$\_$enable signal will never be low when the bus propagates to the interface. 

IBM has an internal tool, a preliminary version of which was presented in \cite{Arbel:2014:ADV}, that is able to identify protected logic along with relevant gating, 
using mix of functional, structural and formal analysis.  The tool reports which latches are protected under which gating conditions, but it does not verify that the gating conditions are proper 
and that there is no {\em over gating} or {\em under gating}; detection of over gating given a functional verification environment is presented in \cite{ArbelBHKKM16}.
Thus, given the results of that tool (which is executed anyway), adding the results of OpenSEA provides the following additional insights:
\begin{itemize}

\item {\em Vulnerable latches due to over gating}:
      Latches that are detected as protected by InsightER,
      but whose flipped value propagates to the macro output due to over gating.
      Such latches may cause silent data corruption.

\item {\em Latches causing spurious alarms}:
      Latches for which the alarm rises,
      although their flipped values do not propagate to the macro output.
      Such latches may cause redundant recoveries, impairing the circuit performance.

\end{itemize}

We run OpenSEA's vulnerability and spurious alarms check
on $15$ macros of varying functionality and sizes from 500 to 7k latches. 
All the experiments were finished within 30min.

We ran tests in which all inputs were kept open. 
The advantage of keeping the inputs open is that this way OpenSEA can be executed as soon as there is a compiled RTL, 
and a full functional simulation environment is not required.
Table \ref{tab:vulnerableAndNotProtected} shows for a few macros the number of latches identified by IBM's tool as not-protected and also as vulnerable by OpenSEA.
This is mainly for validation of the new OpenSEA technology---it demonstrates that OpenSEA indeed detects a significant fraction of the not-protected latches. 

\begin{table}[htbp]
\centering
\begin{tabular}{|c|c|c|}
\hline 
%\textbf{\emph{not-protected InsightER}}  & \textbf{\emph{vulnerable by OpenSEA}}  \tabularnewline
%\textbf{\emph{IBM tool}} & \textbf{\emph{and IBM not-protected}} 		 \tabularnewline
\textbf{\emph{not-protected (I)}}  & \textbf{\emph{\specialcellC{not-protected (I) $\land$\\vulnerable (O)}}}  \tabularnewline
\hline
\hline 
 $162$  & $119$   \tabularnewline
 $32$  & $19$  \tabularnewline
 $1851$  & $535$  \tabularnewline
 $279$ & $109$ \tabularnewline
 $133$ & $110$ \tabularnewline
 $807$ & $606$ \tabularnewline
 $274$ &	$111$  \tabularnewline
\hline
\end{tabular}
\caption{The number of latches found vulnerable by OpenSEA (column ``vulnerable (O)'') and not-protected by InsightER (column ``not-protected (I)'')}
\label{tab:vulnerableAndNotProtected}
\end{table}

%From a practical point of view, this table is significant since its content could indicate a bug in the protection mechanism. However, sequential elements which were detected as vulnerable since the protection mechanism is not entirely in scope, e.g.\ the parity generation is out of scope of the current macro are trivial and not interesting; For these cases the tool should be re-executed and analyzed in a larger scope. Thus, the last column omits these cases and presents latches that have gating on the way to the alarm and OpenSEA gives an indication that these gating are over-gating. These cases indicate a potential bug and should be reviewed by the design team, along with the counter example provided by the tool. 

Table \ref{tab:VulnerableAndProtected} presents the number of latches that are identified as protected by IBM's tool but as vulnerable by OpenSEA. 
From a practical point of view, this table is significant since its content could indicate a bug in the circuit protection mechanism.
However, some of the detected cases are actually irrelevant:
cases in which the parity bit was generated externally and passed as input and the vulnerability is due to an mismatching value of this parity bit, which as input was left open. 
For these cases the tool should be re-executed and analyzed in a larger scope;
also, environment assumptions supported by OpenSEA could help to filter out such results.
Thus, the last column omits these cases and presents latches that have gating on the way to the alarm and OpenSEA gives an indication that these gating are over-gating. These cases indicate a potential bug and should be reviewed by the design team, along with the counter example provided by the tool.

\begin{table}[htbp]
\centering
\begin{tabular}{|c|c|c|}
\hline 
\textbf{\emph{protected (I)}}  & \textbf{\emph{\specialcellC{protected (I) $\land$\\vulnerable (O)}}}  & \textbf{\emph{\specialcellC{protected (I) $\land$\\vulnerable (O) $\land$\\non-trivial}}} \tabularnewline
%\textbf{\emph{IBM tool}} & \textbf{\emph{and IBM protected}} 						& \textbf{\emph{and IBM protected}} \tabularnewline
\hline
\hline 
 $202$  & $52$  & $21$  \tabularnewline
 $548$  & $174$  & $110$  \tabularnewline
 $151$  & $54$  & $54$  \tabularnewline
 $834$  & $206$  & $90$  \tabularnewline
 $1932$ & $208$  & $199$  \tabularnewline
 $104$  & $27$  & $4$ \tabularnewline
 $615$	& $80$ & $26$ \tabularnewline
 $1320$	& $133$ & $91$ \tabularnewline
\hline
\end{tabular}
\caption{The number of latches found vulnerable by OpenSEA (written ``vulnerable (O)'') and protected by InsightER (written ``protected (I)'')}
\label{tab:VulnerableAndProtected}
\end{table}

% erhmm, not sure --- do env assumption allow to spy on latch values? I think no... to be implemented:)

%This way the tool can be executed as soon as there is a compiled RTL, and does not require a functional simulation environment. 
%Table \ref{tab:vulnerableAndNotProtected} shows for a few macros the number of latches identified by IBM's tool as not-protected and also as vulnerable by OpenSEA.
%This is mainly relevant for validation of the new OpenSEA technology - it demonstrates that OpenSEA detects a significant fraction of the not-protected latches.

%Table \ref{tab:VulnerableAndProtected} presents the number of latches that are identified as protected by IBM's tool but as vulnerable by OpenSEA.
%From a practical point of view, this table is significant since its content could indicate a bug in the protection mechanism. However, seuqential elements which were detected as vulnerable since the protection mechanism is not entirely in scope, e.g.\ the parity generation is out of scope of the current macro are trivial and not interesting; For these cases the tool should be re-executed and analyzed in a larger scope. Thus, the last column omits these cases and presents latches that have gating on the way to the alarm and OpenSEA gives an indication that these gating are over-gating. These cases indicate a potential bug and should be reviewed by the design team, along with the counter example provided by the tool. 

In the second batch of experiments we used OpenSEA to find spurious alarms. Namely, cases in which the alarm is being raised even though the output is not being affected. %
The results are in Table~\ref{tab:falseAlarmOpenSEA}.
At first, we found many spurious alarms.
But many of them were irrelevant due to injecting faults into protection circuit itself.
OpenSEA allows one to configure latches into which to inject faults.
When we excluded the latches of the protection circuit, the number of spurious alarms dropped.
Each case left is a potential performance/power issue which may indicate that the existing gating is not tight enough and should be improved.
This indication is not provided by IBM's internal tool or by any other tool we are familiar with.

\begin{table}[htbp]
\centering
\begin{tabular}{|c|c|c|}
\hline 
\textbf{\emph{protected (I)}}  & \textbf{\emph{spurious alarm (O)}}  & \textbf{\emph{\specialcellC{spurious alarm (O)\\excl. protection logic}}} \tabularnewline
%\textbf{\emph{IBM tool}} & \textbf{\emph{OpenSEA}} 						& \textbf{\emph{not checking logic}} \tabularnewline
\hline
\hline
 $202$  & $86$  & $57$  \tabularnewline
 $660$  & $244$  & $122$  \tabularnewline
 $1932$  & $4$  & $4$  \tabularnewline
 $2104$ & $3$ & $0$ \tabularnewline
 $439$ & $51$	& $28$ \tabularnewline
\hline
\end{tabular}\caption{The number of latches whose faults can cause spurious alarms found by OpenSEA (written ``spurious alarm (O)'')
 and protected by InsightER (written ``protected (I)'').
 For spurious alarms, we run OpenSEA only on the latches found to be protected by InsightER.
 }
\label{tab:falseAlarmOpenSEA}
\end{table}

Finally,
Table~\ref{tab:timeConsumption} presents the time consumption of OpenSEA's search algorithms for vulnerable latches and for spurious alarms.
Obviously, time consumption is very low.

\begin{table}[H]
\centering{}\begin{tabular}{|c|c|c|}
\hline 
\textbf{\emph{\# latches}}       & \textbf{\emph{\specialcellC{Time: searching\\vulnerable latches}}}  & \textbf{\emph{\specialcellC{Time: searching\\spurious alarms}}} \tabularnewline
%\textbf{\emph{latches}} & \textbf{\emph{vulnerability }} & \textbf{\emph{false fail}} 	\tabularnewline
\hline
\hline
 $434$  & $7.45$ sec.  & $<\!1$ sec.  \tabularnewline
 $582$  & $9.65$ sec.  & $<\!1$ sec.  \tabularnewline
 $590$  & $9$ sec.  & $1.84$ sec.  \tabularnewline
 $7061$ & $22$ min. $23$ sec. & $2$ min. $31$ sec. \tabularnewline
\hline
\end{tabular}\caption{Timings of OpenSEA algorithms}
\label{tab:timeConsumption}
\end{table}

\section{Related Work}

Fey, Frehse and Drechsler~\cite{bmc_approx} classify components into
(i) vulnerable,
(ii) protected,
(iii) dangerous, and
(iv) unknown.
To this end, they use a bounded model-checking based technique to compute the circuit robustness,
which is the ratio of the protected latches to all the latches.
Furthermore, by under- and over- approximating the reachable states
they can compute upper and lower bounds on the robustness.
Our approach for computing vulnerable and protected latches is very similar to theirs.
The differences are:
(i) we can use real test cases, while they consider only symbolic test cases,
(ii) moreover, we study the question of false alarms,
(iii) furthermore, we introduce the notion of ``environment model'' which is handy for checking practical designs,
(iv) in contrast, they study ATPG-based algorithms and consider optimization ``dominator''.

A year later, the same three authors together with Arbel and Yorav in~\cite{DBLP:conf/fmcad/FrehseFAYD12}
proposed a more efficient approach to classify the components.
This time, they use interpolation for an over-approximation of the reachable state space and compute fixed-points.
This way, only states that are relevant to the property to prove are considered.
\ak{more details here}

In 2006, Krautz et al.~\cite{Krautz} proposed a way to evaluate the coverage of error detection logic using BDDs.
They used a fault injection model similar to our BMC based approach in which they compare
a fault-free device with a faulty one.
The output of their fault injection model is defined by a property checker that defines 
multiple properties by comparing primary outputs of both circuit copies.
A sequential equivalence check of this circuit is performed up to a defined number of time steps by creating a BDD representation.
The number of input combinations that make properties true are counted.
Among other properties, they count a property similar to our definition of \emph{vulnerabilities} and the number of injected faults.
A coverage is computed using these numbers.

Holcomb et al.~\cite{Holcomb} showed another way to compute the \emph{failure in time}
rate by performing a system-level analysis.
In contrast to our algorithms, their methods rely on formal specifications.
Our approaches only need to know which output is the \emph{alarm} output.

In 2014, Arbel, Koyfman, Kudva and Moran published a structural approach to detect parity-protected memory elements~\cite{Arbel:2014:ADV}.
For this purpose, they first search for potential error detection circuits and for latches that are potentially protected by those using functional analysis.
Afterwards, they use SAT-calls to prove that these latches are indeed protected by the parity net. The unique characteristic of their method is that they
analyze a circuit locally rather than looking at the behavior of the entire system.
Their method can be applied whenever error checker latches are present,
which is not always the case though.

Finally, methods that systematically construct robust systems\cite{roco}
might make our work and all methods to verify error detection and error correction obsolete one day.
However, as long as these intelligent compilers are not yet perfect (if they will ever be) verification of circuits will stay important.

\section{Conclusion}
In this paper,
we presented soft-error analysis framework OpenSEA to find vulnerable latches, protected latches, and to find spurious alarms.
OpenSEA supports many variations of the algorithms---SAT-based, BDD-based, and enumerative---and many SAT solvers as a back end.
The algorithms can use symbolic, semi-symbolic, and concrete tests.
In the experiments, we run OpenSEA on industrial macros from IBM
to search for spurious alarms, vulnerable, and protected latches.
We explained why not all the bugs found are real and described the OpenSEA features to avoid such spurious bugs.
The source code, benchmarks, and documentation can be found at
\url{https://extgit.iaik.tugraz.at/scos/soft-error-analysis/}.

\section*{Acknowledgment}
\small{
This work was supported by the European Commission through 
        the project IMMORTAL (644905) and by the Austrian Science Fund
        (FWF) through the research network RiSE (S11406-N23).}

\bibliographystyle{IEEEtran}
\bibliography{references}
\end{document}